\documentclass[journal]{vgtc}                     

\onlineid{1009}

\vgtccategory{Research}

\vgtcpapertype{Theoretical \& Empirical}

\title{Affective Visualization Design: \\Leveraging the \x{Emotional Impact} of Data}

\author{Xingyu Lan,Yanqiu Wu, and Nan Cao}

\authorfooter{
\item
Xingyu Lan is with Fudan University and a member of the Research Group of Computational and AI Communication at Institute for Global Communications and Integrated Media; 
Yanqiu Wu and Nan Cao are with
Intelligent Big Data Visualization Lab at Tongji University.\\
E-mail: \{xingyulan96, wuyanqiu.idvx, nan.cao\}@gmail.com. Nan Cao is the corresponding author.
}

\abstract{%
  In recent years, more and more researchers have reflected on the undervaluation of emotion in data visualization and highlighted the importance of considering human emotion in visualization design. Meanwhile, an increasing number of studies have been conducted to explore emotion-related factors. However, so far, this research area is still in its early stages and faces a set of challenges, such as the unclear definition of key concepts, the insufficient justification of why emotion is important in visualization design, and the lack of characterization of the design space of affective visualization design. To address these challenges, first, we conducted a literature review and identified three research lines that examined both emotion and data visualization. We clarified the differences between these research lines and kept 109 papers that studied or discussed how data visualization communicates and influences emotion.
  Then, we coded the 109 papers in terms of how they justified the legitimacy of considering emotion in visualization design (\ie \textit{why} emotion is important) and identified five argumentative perspectives. Based on these papers, we also identified 61 projects that practiced affective visualization design. We coded these design projects in three dimensions, including design fields (\textit{where}), design tasks (\textit{what}), and design methods (\textit{how}), to explore the design space of affective visualization design.
}


\graphicspath{{figs/}{figures/}{pictures/}{images/}{./}} 

\usepackage{tabu}                      
\usepackage{booktabs}                  
\usepackage{lipsum}                    
\usepackage{mwe}                       

\newcommand{\etal}{et~al.~} 
\newcommand{\ie}{i.e.,~}
\newcommand{\eg}{e.g.,~}

\usepackage{microtype}                 
\PassOptionsToPackage{warn}{textcomp}  
\usepackage{textcomp}                  
\usepackage{mathptmx}                  
\usepackage{times}                     
\usepackage{cite}                      
\usepackage{tabu}                      
\usepackage{booktabs}                  
\usepackage{enumitem}
\usepackage{url}
\usepackage{soul}
\makeatletter
\g@addto@macro{\UrlBreaks}{\UrlOrds}
\makeatother
\usepackage{graphicx}
\usepackage{wrapfig}
\usepackage[T1]{fontenc}
\usepackage{amsmath}  
\usepackage{multirow}
\usepackage{tabularx}
\newcommand{\x}[1]{\textcolor{black}{#1}}
\usepackage[svgnames,x11names,table]{xcolor}

\usepackage{mathptmx}                  

\begin{document}

\firstsection{Introduction}
\maketitle

In the 1880s, philosopher Friedrich Nietzsche wrote down his eloquent argument, "thoughts are the shadows of our feelings"~\cite{nietzsche2022gay}. Although scientists are still investigating the role and mechanism of emotion, there is a general consensus that emotion is an indispensable part of human intelligence and has a crucial influence on perception, cognition, and behavior~\cite{lewis2010handbook}.
In the visualization community, although traditional criteria of evaluating data visualization design emphasize rational metrics such as accuracy and efficiency, an increasing amount of work has recognized the importance of emotion in visualization~\cite{d2020data,wang2019emotional,lan2021smile}. 
For example, empirically, emotion has been observed as a commonly pursued factor in fields such as artistic visualization~\cite{kosara2007visualization,viegas2004artifacts,wang2019emotional} and data storytelling~\cite{lan2021kineticharts,lan2022negative,shi2021communicating,yang2021design,shi2022breaking}, capable of influencing user engagement and the recall of data~\cite{kennedy2018feeling, lan2022negative}.
Theoretically, researchers have introduced concepts such as \textit{emotional design} and \textit{affective design} from domains such as industrial design and human-computer interaction (HCI) to help examine the relationship between data visualization and emotion~\cite{lan2021kineticharts,lan2021smile,bartram2017affective}. 
Meanwhile, many data visualization projects that intend to leverage the benefits of emotion have been proposed.
For instance, researchers have explored affective data visualization as a promising approach to communicating data to a wide audience~\cite{aragon2021risingemotions,perovich2020chemicals}.
In the wild, design projects such as the U.S. Gun Death~\cite{gun} have sparked heated discussion on the role of visualization in evoking emotion and doing good for society~\cite{boy2017showing}.

However, the development of this thread of research still faces three main challenges. 
Firstly, there is a need for a clear definition and characterization of the research scope.
For example, the concept, \textit{affective visualization}, has been used inconsistently in different papers (\eg ~\cite{bartram2017affective,zhang2010affective}), thus hindering effective communication between researchers.
Secondly, there is still controversy surrounding the legitimacy of considering emotion in visualization design.
For example, from the perspective of traditional rationalism-driven design norms, emotion seems to be at odds with reasoning. Therefore, a frequently raised doubt is whether affective design will hinder the objectivity of data~\cite{lee2022affective,d2020data}.
However, to our best knowledge, no work has been done to systematically review the relevant arguments that are now scattered in various publications and legitimize \textit{why} emotion is important for visualization design.
Thirdly, we still lack a clear picture of how to operationalize affective design on visualization, such as \textit{where} to apply such design, \textit{what} tasks the design can undertake, and \textit{how} to perform the design. 
Only by answering such questions can we sort out existing knowledge and practices in this growing field and suggest concrete strategies and techniques for designers.

To address the above challenges, first, we conducted a literature review and identified three research threads that examined both emotion and data visualization. We clarified the differences between these research threads and set one specific thread, \textit{affective visualization design} (data visualization designed to communicate and influence emotion), as the focus of this work (\cref{sec:corpus}).
Next, we coded 109 papers concerning affective visualization design based on how they argued for the legitimacy of considering emotion in visualization design (in response to the \textit{why} challenge mentioned above) and categorized the arguments into five perspectives, including the perspectives of application, usefulness, rhetoric, \x{sociology}, and humanism (\cref{sec:argument}).
Based on the 109 papers, we also identified 61 affective visualization design projects whose authors have explicitly stated their intent to communicate and influence emotion. We coded these design projects in three dimensions (\ie design fields, design tasks, design methods) corresponding to the \textit{where}, \textit{what}, and \textit{how} challenges to explore the landscape of affective visualization design (\cref{sec:space}). Last, we discuss our limitations and the research opportunities arising from this work (\cref{sec:discussion}).

\section{Background}
In this section, we review background knowledge about affective design and the debate about emotion in visualization design.

\subsection{Affective Design}

Affect, as defined in the APA Dictionary of Psychology~\cite{apa}, represents any experience of feeling or emotion, and is a critical component of human intelligence (for simplicity, we follow prior work~\cite{picard2000affective} and use \textit{affect} and \textit{emotion} in the following texts interchangeably). 
Over the years, researchers and practitioners have explored the mechanism and benefits of emotion and applied the knowledge to various designs.
For example, in the 1980s and 1990s, a significant amount of psychological research was done to investigate the relationship between color and emotion~\cite{valdez1994effects}. 
Today, color, along with other emotional elements such as shapes, fonts, and iconography, are widely used in graphic design to create emotive posters or web pages~\cite{walter2011designing}.
In the field of industrial design, 
researchers started exploring \textit{emotional design}~\cite{norman2004emotional} in the early 2000s, focusing on eliciting emotion by manipulating the appearances and functions of industrial products such as cars and furniture. 
In the field of HCI, designing for emotion gained popularity especially after the conceptualization of
\textit{affective computing}~\cite{picard2000affective}. Ideas such as \textit{affective design} and \textit{affective user interface} were proposed and embraced by HCI researchers to develop affective chatbots, computer games, and VR envionments~\cite{beale2008role,pinilla2021affective,johnson2003effective}. 

In the visualization community, it is not until recent years that researchers began to highlight the opportunities of incorporating affective design in data visualization~\cite{bartram2017affective,lan2021smile,lan2021kineticharts}. Following the call, this work conducts a systematic review of existing research and practices on affective visualization design and outline the latest advances in this emerging cross-disciplinary field.

\subsection{The Debate About Emotion in Visualization Design}
\label{sec:debate}

Although affective design is a well-recognized design paradigm in many fields, it remains underexplored in the visualization community~\cite{lan2021smile}.
As mathematical and factual depictions of information, data does not seem directly related to emotion.
The mainstream visualization design norms highlight that a good visualization should be accurate, concise, and clear~\cite{tufte2001visual}. Under such a rationalism and minimalism design paradigm, the legitimacy of including emotional elements in data visualization faces challenges. For example, in the debate about chart junk, the use of emotion-laden embellishments was criticized as redundant and distracting~\cite{tufte2001visual}.
Another common concern is that including emotion may reduce the objectivity of data. 
According to historian Porter~\cite{porter1996trust}, the pursuit of objectivity is a core spirit of modern science, and quantification is viewed as "a technology of distance", where people should separate their subjective emotions from objective facts.

On the other hand, different opinions constantly exist. As early as 1995, a panel discussion was held at the IEEE Visualization Conference to reflect on "Is visualization struggling under the myth of objectivity?"~\cite{jorgenson1995is}. In this panel, researchers discussed the tension between viewing visualization as objective science and the inherent subjectivity of visual design. One researcher argued that "visualization is resistant to the systematic evaluation and assessment procedures common to science; it still remains difficult to ascertain if a particular instantiation has been 'successful'".
In recent years, more and more arguments, reflections, and critiques have been proposed.
For example, Dragga~\cite{dragga2001cruel} responded to the chart junk debate by arguing that although pictographs are statistically redundant, they are not emotionally redundant.
Kennedy~\etal~\cite{kennedy2016work} thought that the binary view of emotion and reason has viewed emotion as irrational and has made emotion intentionally undervalued in data science for a long time. In the wild, many practitioners have created emotion-laden visualizations and defended the necessity of communicating emotion with data~\cite{lan2022negative,boy2017showing}.
However, so far, such arguments and practices remain highly scattered and have not been systematically reviewed by the visualization community.
For example, no systematic work has been done to clarify why it is important to consider emotion in visualization design. Additionally, no summarization has been done to clarify what industries or needs can be served by affective visualization design and how it should be designed.
To fill these gaps, this work aims to synthesize relevant research work and design practices about affective visualization design to clarify the values of emotion in visualization and derive design implications.

\section{Corpus Collection}
\label{sec:corpus}
This section reviews the literature on affective visualization design. Below we first introduce how we collected and refined the corpus, and then provide an overview of the corpus.

\subsection{Methodology}

To start with, we \x{went through papers published in three well-recognized leading venues of data visualization research (IEEE TVCG, ACM CHI, and EuroVis)} in the recent three years (2020-2022) to identify an initial set of qualified papers. As prior literature has suggested that affective visualization design is close to art practice~\cite{wang2019emotional,kosara2007visualization}, we also \x{went through} papers published in IEEE VIS Arts Programs in the recent three years. 
Specifically, we considered a paper qualified if: (i) it is about data visualization, and (ii) it explicitly mentions emotion or emotion-related words (\eg surprise, joy) in its keywords, title, abstract, or major findings.
As a result, we identified 24 qualified papers~\cite{lan2021smile,lan2021kineticharts,lan2022negative,lan2021understanding,lee2022affective,concannon2020brooke,perovich2020chemicals,romat2020dear,anderson2021affective,sallam2022towards,rebelo2022essys,zeller2022scientific,aseniero2022skyglyphs,samsel2021affective,elli2020tied,qin2020heartbees,padilla2022multiple,morais2022exploring,liem2020structure,lee2020data,floyd2021eat,morais2021can,ajani2021declutter,kauer2021public}.

For example, 
Samsel~\etal~\cite{samsel2021affective} designed a color tool for creating affective scientific visualization. 
Lan~\etal~\cite{lan2022negative} investigated how visualization design facilitates the communication of negative emotions in serious data stories.
\x{Based on the 24 papers}, we used the backward snowballing method (\ie going through the reference lists of existing papers to identify new papers to include) to identify more \x{potentially qualified papers and used the aforementioned two criteria to judge whether a paper is truly qualified. For example, some papers about user experience were deemed as potentially relevant during the snowball process. However, a further search in their texts showed that emotion is not a focal point of their research. Therefore, these papers were not included in our corpus.} If two papers have highly overlapping content (\eg a short paper and a full paper written by the same authors and dealt with the same research problem), we kept the paper whose research is more comprehensive.
We iterated the above process until no more qualified papers were identified.

\subsection{Concept Clarification and Corpus Refinement}

When constructing the corpus, a challenge that emerged immediately was the ambiguity of research boundaries and concept usage. For example, although some papers have indeed studied both visualization and emotion and were tagged as qualified in the first cycle of paper collection, their research goals are different.
For example, we noticed that \textit{affective visualization} has been used in different contexts to refer to different meanings. 
To address this challenge and also clarify the research scope of this work, we categorized the papers into three research threads to elucidate their differences and refine our corpus.

\textbf{Emotion as the object being visualized.}
This thread of research performs visual analytics on the data that reflects emotion, such as heart rates, facial expressions, and gestures. In other words, emotion is the object being visualized and analyzed. For example, Zhang~\etal~\cite{zhang2010affective} developed a visual analytics system to help mine the emotion in music videos and called the visualization in the system \textit{affective visualization}. Recently, when studying virtual reality, a group of researchers used \textit{affective visualization} to describe the visualization of people's affective states in immersive environments~\cite{pinilla2021affective}. Other similar work includes visualizing the emotion in presentations and TED talks ~\cite{zeng2019emoco,wang2021dehumor}.

\textbf{Emotion as a precondition of viewing visualization.}
This thread of research views emotion as a precondition that exists before viewing visualization. For example, Harrison~\etal~\cite{harrison2013influencing} studied the priming effect of emotion by first using texts to evoke positive or negative emotions from the participants, then asking the participants to view charts, and examining whether the emotions would influence the accuracy of graphical perception. 
Thoresen~\etal~\cite{thoresen2016not} studied whether people's anxiety levels would influence their modes of reading maps.



\textbf{Emotion as a result of visualization design.}
This thread of research studies emotion as a result of viewing data visualization and highlights the need of considering viewers' subjective experiences in visualization design. 
For example, Bartram~\etal~\cite{bartram2017affective} defined \textit{affective visualization} as a type of visualization that "uses visual features to evoke a mood, feeling or impression" and proposed a set of color palettes that can effectively convey emotion in chart design. Lan~\etal~\cite{lan2021kineticharts} presented Kineticharts, an animation design scheme for creating animated charts that express five emotions commonly desired by data storytelling.


In summary, we identified multiple threads of research that stand at the intersection of visualization and emotion. 
In pursuit of a better conceptualization of these research threads and facilitate clearer dialogues between researchers, we propose: (i) As the concept, \textit{affective visualization}, has been used in a broad sense and has already involved diverse work about emotion, its definition can be more inclusive and encompassing. By referring to the definition of 
\textit{affective computing}~\cite{picard2000affective}, we define \textit{affective visualization} as a data visualization that relates to, arises from, or influences emotion. 
(ii) What this work intends to focus on is the third research thread (\ie emotion as a result of visualization design). As the core issue that distinguishes this research thread from the other two is its emphasis on the role of design, we use the term \textit{affective visualization design} in the following texts to refer to a class of data visualizations designed to communicate and influence emotion.

Next, we refined our corpus and only kept papers about affective visualization design. 
Note that we also found four papers that addressed hybrid research problems. For example, Qin~\etal~\cite{qin2020heartbees} proposed a bio-feedback system called HeartBees to visualize the emotion of the crowd. Meanwhile, the authors also intentionally used metaphorical shapes and animation to create an affective visual representation. As they also contributed knowledge about affective visualization design, we included the four papers in our corpus. The final corpus contains 109 papers in total (see \url{https://affectivevis.github.io/}).

\subsection{Corpus Overview}

In general, the number of papers concerning affective visualization design has grown remarkably over the years. Before 2006, at most one paper was published every year, but in 2022 alone, 14 papers were published.
We then analyzed which research areas the papers belong to. We searched the publication venues of these papers on the Web of Science (WoS) and identified the research areas tagged to these venues. For example, the research area of IEEE TVCG and the proceedings of ACM CHI was tagged as \textit{Computer Science} on WoS.
For 17 papers that were not searchable on WoS, we searched the home pages of their venues manually and coded their research areas based on how these venues introduced and defined themselves using WoS's taxonomy~\cite{wos}.
Two authors first coded independently and then compared their codes (Cohen's Kappa = 0.82) and discussed until achieving 100\% agreement. 
For venues that have multiple area tags, we used their primary areas to perform the following statistics.

\begin{figure}[!b]
 \centering
 \vspace{-1em}
 \includegraphics[width=\columnwidth]{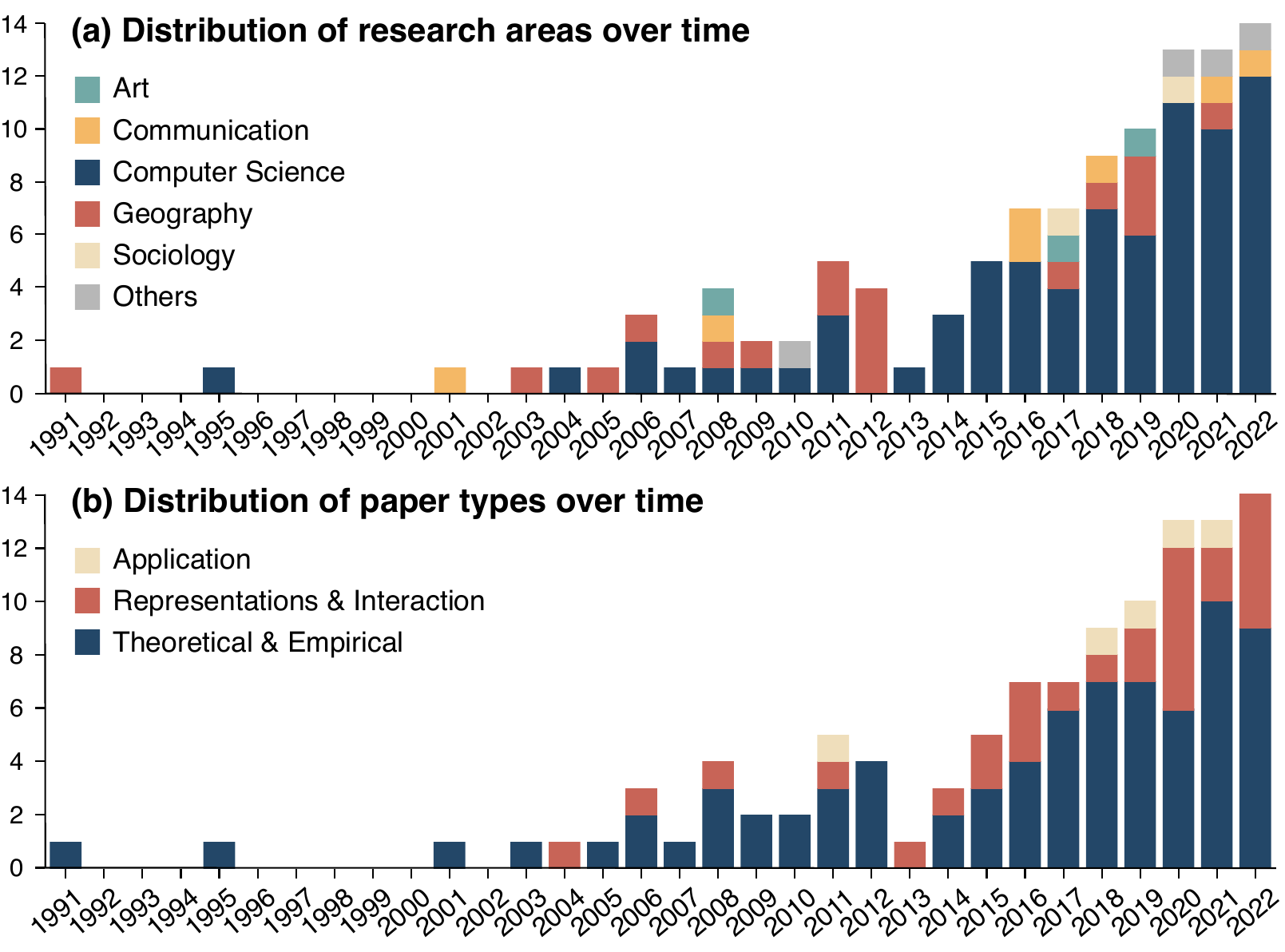}
 \caption{The distribution of the 109 papers in our corpus over time in terms of their (a) research areas and (b) paper types.}
 \label{fig:corpus}
\end{figure}

Firstly, we counted the number of papers published in each research area and identified five areas that have more than one paper concerning affective visualization design, including \textit{Computer Science} (75), \textit{Geography} (18), \textit{Communication} (7), \textit{Art} (3), and \textit{Sociology} (2). Four papers were published in other areas, such as \textit{Chemistry} (1) and \textit{Thermodynamics} (1).
In other words, we found that affective visualization design has attracted the interest of researchers from multiple communities.
In terms of temporal trends, as shown in \cref{fig:corpus} (a), the community of \textit{Geography} has shown earlier interest in this research topic. For example, in 1991, Gilmartin and Lloyd~\cite{gilmartin1991effects} conducted a user study to examine how different map projections influence people's emotional involvement with places.
In the early 2010s, a group of geographers proposed ideas such as \textit{affective cartography}~\cite{iturrioz2011artistic} and \textit{affective geovisualization}~\cite{aitken2011affective}, which pushed this line of research to its peak.
It is also in the 2010s that the papers published in the area of \textit{Computer Science} began to mushroom. To date, papers belonging to this area have become the main part of affective visualization design research.

Next, we coded the research types of these papers following the six categories (\ie \textit{Theoretical \& Empirical}, \textit{Applications}, \textit{Systems \& Rendering}, \textit{Representations \& Interaction}, \textit{Data Transformations}, \textit{Analytics \& Decisions}) defined by the IEEE VIS conference~\cite{area}.
As a result, as shown in \cref{fig:corpus} (b), most papers belong to the \textit{Theoretical \& Empirical} category (76). These papers either proposed theoretical frameworks, critiques, and observations or examined the relationship between visualization design and emotion by conducting user studies.
The second most identified paper type is \textit{Representations \& Interaction} (28), which explored or proposed new visualization approaches to influencing emotion. 
A small number of papers belong to the \textit{Applications} category (5), which built visualization systems or tools to help generate or create affective visualization design.
Chronologically, \textit{Theoretical \& Empirical} has also been the dominant research type. However, in recent years, we see more novel methods (\textit{Representations \& Interaction}) and design tools (\textit{Applications}) being proposed to enrich the toolkit of affective visualization design and lower its authoring barrier.

\section{Why Affective Visualization Design}
\label{sec:argument}
As mentioned in \cref{sec:debate}, as a design paradigm that embraces subjective experiences, affective visualization design still faces controversy. 
To address the controversy, we coded the 109 papers in our corpus following the methodology of thematic analysis~\cite{braun2012thematic} with the goal of distilling existing arguments in support of affective visualization design.
Two authors first read the papers one by one and marked any sentences that explain the authors' justifications for studying affective visualization design. 
Then, we went through these sentences and generated codes independently to summarize the rationales behind the arguments. To further identify different perspectives of argumentation, we also grouped similar codes into high-level themes.
For example, Kosara~\cite{kosara2007visualization} argued that emotion is important for the domain of data art, because artists usually value the sublime quality of design more than the pragmatic quality.
Therefore, this argument was coded as "emotion is prioritized in some domains" and was assigned to a theme called "from the perspective of application".
Then, we met to compare our codes, unify the wording of similar codes, and resolve disagreements through iterative discussion. 
Last, we tagged the papers again using the final coding scheme. In total, we identified \x{seven} different arguments categorized into five perspectives. 

\begin{figure*}[t]
 \centering
 \includegraphics[width=\textwidth]{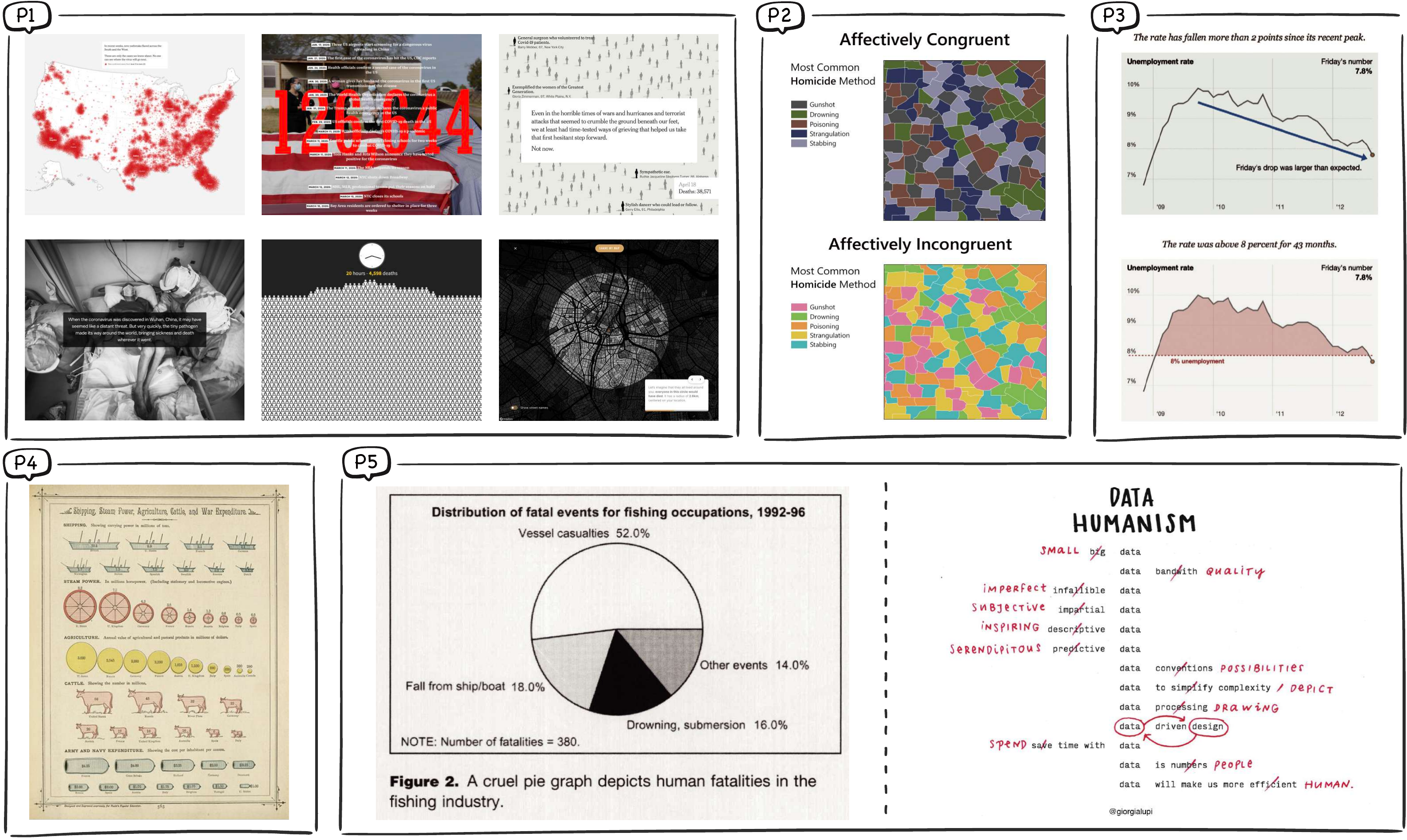}
 \vspace{-1.5em}
 \caption{Examples of the five argumentative perspectives: (P1) affective data stories published by news media during the COVID-19 pandemic~\cite{lan2022negative}, (P2) emotional color palettes can facilitate the cognitive interpretation of maps~\cite{anderson2021affective}, (P3) two objective-looking but rhetorical charts~\cite{d2020data}, (P4) a vintage pictograph that practiced Victorian values~\cite{kostelnick2016re}, (P5) a "cruel pie" about human fatalities~\cite{dragga2001cruel} and the call for data humanism~\cite{lupi}.}
 \label{fig:arguments}
 \vspace{-1.5em}
\end{figure*}

\subsection{From the Perspective of Application}

As data visualization is serving a wide range of domains and users, the design requirements of data visualization are becoming increasingly diverse. In our corpus, 76 papers argued the legitimacy of affective visualization design from the perspective of realistic needs.

\textbf{Emotion is valued by certain domains.}
Different domains have different needs, values, and priorities. For example, for data artists, emotion is a vital channel of self-expression and can spark imagination and creativity~\cite{claes2017public,wang2019emotional}.
For the media industry, communicating emotion through visualization design is also common, and multiple studies have observed the integration of affective visualization design in various visual stories published by news agencies, governments, or non-profits~\cite{lan2021smile,lan2022chart}.
For example, during the COVID-19 pandemic, many data stories were created to evoke shock, fear, or grief for deaths~\cite{lan2022negative} (\cref{fig:arguments} P1).
In addition, emotion is also valued by domains such as education and entertainment~\cite{gough2014affective,carpendale2017subjectivity}.
For example, when investigating the popularity of Wordle (a web-based tool for visualizing texts), Viégas~\etal~\cite{viegas2009participatory} concluded that "emotional resonance seems to be a key reason for preference of wordles over tag clouds" because it enables the public to play with data and customize their own visualizations.

\textbf{New application scenarios are emerging.}
Several researchers justified the necessity of considering emotion in data visualization design based on new technology trends and market needs. For example, in 2004, given the trend that "the contents of digital archives that permeate our daily lives become more emotionally charged", a set of systems that visualize personal data were developed~\cite{viegas2004artifacts}.
Around 2010, Cartwright~\etal~\cite{cartwright2008developing} and van Lammeren~\etal~\cite{van2010affective} proposed that the maturity of Web 2.0 technology and 3D modeling technology can expand the functions of maps. For example, people can now use maps to fulfill more affective intents, such as annotating places with their cherished stories or sharing the stories with beloved ones.

\subsection{From the Perspective of Usefulness}
Secondly, \x{67} papers justified the role of emotion in visualization design by providing evidence from empirical studies or neuroscience to demonstrate emotion is useful and can bring benefits.

\textbf{Emotion is not the enemy of rationality.}
In response to the challenge that emotion may hinder rational thinking about data, Kennedy and Hill~\cite{kennedy2018feeling} cited various arguments from sociologists and philosophers to criticize the stereotype of the "reason/emotion binary".
Aitken and Craine~\cite{craine2011emotional,aitken2011affective} argued for the legitimacy of conveying emotion in cartography by citing work from neuroscience, which has proven that emotion is not separate from, but is a huge part of rationality.
When studying the affective design of infographics, Lan~\etal~\cite{lan2021smile} referred to the dual-process theory in psychology to demonstrate that emotion is heavily involved in thought processes and can co-exist with rationality.
Similarly, Kostelnick~\cite{kostelnick2016re} and D'ignazio~\etal~\cite{d2020data} agreed that emotion can complement rather than contradict the logical analysis of data. 
In several user studies~\cite{lan2022negative,lan2021kineticharts}, researchers indeed observed that affective visualization design was not at odds with data comprehension.
Anderson and Robinson~\cite{anderson2021affective} found that maps whose color palettes convey congruent affective messages with content were felt more appropriate and less confusing than those \x{that} use incongruent colors, showing that emotion can work with or even amplify rationality (\cref{fig:arguments} P2).

\textbf{Emotion brings additional benefits.}
First, emotion has been found useful for enhancing user engagement with data. For example, Lan~\etal~\cite{lan2021kineticharts} found that in data storytelling, emotion showed a significantly positive correlation with engagement-related metrics such as focused attention, enjoyment, interest, and likability. Qualitatively, Kennedy and Hill~\cite{kennedy2018feeling} observed that people were more likely to spend time with a data visualization when being emotionally aroused by its design or content. 
In addition, emotion has also been observed to enhance people's perceived connection with data~\cite{gilmartin1991effects,lan2021kineticharts} and influence how they memorize data~\cite{lan2022negative}.
Emotion also influences people on a deeper level, such as changing attitudes, values, or behaviors.
For example, Boy~\etal~\cite{boy2017showing} and Morais~\etal~\cite{morais2021can} cited prior work from psychology and sociology to illustrate that emotions such as empathy can help people perceive others' misery and promote prosocial behavior.
Lan~\etal~\cite{lan2022negative} used a set of serious data stories that convey negative messages about COVID-19 as stimuli and found that appropriate design methods helped evoke deep emotions and strengthened contemplative thoughts and self-reflection.

\subsection{From the Perspective of Rhetoric}
\x{Seventeen} papers argue that data visualization design is a rhetorical device that always integrates subjective factors.

\textbf{Objective data visualization does not exist.}
In response to the challenge that adding emotion to visualization may make data less objective, many researchers argued that no data visualization is truly objective, because data visualization design, as all design forms, is rhetorical~\cite{campbell2018feeling,muehlenhaus2012if,dragga2001cruel,lee2022affective,d2020data,kennedy2016work}.
For example, D'ignazio and Klein~\cite{d2020data} took two charts posted on New York Times as examples (\cref{fig:arguments} P3); the two charts both look concise, "objective", and are high in the data-ink ratio, but they deliver different messages (the blue chart shows the unemployment rate in the US decreased during the Democrats' administration while the red chart shows the rate is constantly high from a Republican perspective). D'ignazio and Klein discussed the rhetorical intentions behind these designs and argued that charts that create the "illusion of objectivity" can be even more deceptive than affective charts. 
Similarly, Kennedy~\etal~\cite{kennedy2016work} thought the use of clean layout and geometric shapes are common conventions that create "the aura of objectivity" in data visualization.
Recently, Lee-Robins and Adar~\cite{lee2022affective} also argued that data is not neutral.

\subsection{From the Perspective of \x{Sociology}}
\x{The arguments of nine papers emphasize that the current tendency to avoid emotion in visualization design is not a golden rule, but is constructed by the historical context, culture, and social values}.

\textbf{\x{The emphasis or suppression of emotion is a social phenomenon.}}
\x{Emotion was once highlighted in history. For example, 
Kostelnick~\cite{kostelnick2016re} found that emotion played an important role in chart design in the 1800s when Romantic values \x{(which emphasized individualism, emotion, and imagination)} dominated all forms of design. At that age, visualization designers were also bold to use lavish color, pictorial elements, or epic narratives to enhance emotional appeal (\eg \cref{fig:arguments} P4).
However, after the Enlightenment, rationalism has become a dominating value, breading the tendency to undervalue emotion~\cite{kostelnick2016re,kennedy2018feeling}. 
In a study, van Koningsbruggen and Hornecker~\cite{van2021just} found that participants described feeling not being allowed to emotionally connect to visualizations and emotion was intentionally got rid of, and they argued that "a form of post-hoc rationalisation takes place, which obscures people’s initial connections and affective responses to visualisations".}

\subsection{From the Perspective of Humanism}
Five papers argue from the angle of humanism, which claims that data visualization should be humane, ethical, and do good to society.

\textbf{Data should not be cold and cruel.}
For example, Dragga~\cite{dragga2001cruel} pointed out that statistical charts designed in a minimalistic style can be emotionless and ignore the real people behind the data.
He took several charts about fatalities and disasters as examples, criticizing them as "cruel bars" and "cruel pies" (\cref{fig:arguments} P5, left). 
Likewise, Mccleary~\cite{mccleary2003beyond} thought that a conventional map design would appear indifferent if it visualizes data about genocide.
In recent years, some researchers~\cite{alamalhodaei2020humanizing,campbell2018feeling,romat2020dear} have recognized or applied the idea of \textit{data humanism} proposed by visualization designer Giorgia Lupi~\cite{lupi} (\cref{fig:arguments} P5, right). In Lupi's original essay, she called, "to envision ways to use data to feel more empathic, to connect with ourselves and others at a deeper level...we have to bring data to life - human life."



\section{Exploring the Landscape of Affective Visualization Design: Where, What, and How}
\label{sec:space}

In this section, we analyze design projects that have practiced affective visualization design and explore what the design space is.

\subsection{Methodology}

Based on the \x{109 papers} we collected earlier, first, we went through all papers and kept papers that are affective visualization design projects themselves (\ie design studies). \x{As a result, 36 out of 109 papers belong to this situation.}
For example, 
Concannon~\etal~\cite{concannon2020brooke} designed a data film to enhance the emotional closeness between people and data.
Lan~\etal~\cite{lan2021kineticharts} proposed a design scheme called Kineticharts for creating animated charts that express emotions. 
Next, we also searched \x{the reference lists of the 109 papers to identify more in-the-wild design projects that are potentially qualified}. 
For example, U.S. Gun Death~\cite{gun} is a design project cited by multiple papers~\cite{lan2022negative,lee2022affective} as a representative example of affective visualization design. 
However, although such design projects were considered affective by researchers, we lacked first-hand justifications and explanations from their authors, thus preventing us from understanding their true design intents or tasks.
Therefore, for each of the design projects, we searched its publication page, the author's personal website, or any interviews or talks that introduced the project to double-check the design intents. For example, we found that the designers of U.S. Gun Death have indeed recognized their affective intents by saying that "it’s a massive collection of human emotion hidden within rows and rows of numbers"~\cite{gun2} and "we need people to react, we need people to sort of get riled up about things, get excited about things, and want to make a change in the world"~\cite{gun3}.
Therefore, we viewed it as a qualified project of affective visualization design.
After the double-checking process, we identified 25 in-the-wild design projects whose designers have explicitly stated their intent to communicate and influence emotion using data visualization. Combining with the 36 qualified projects identified earlier, we in total collected 61 projects as the exemplars of affective visualization design. 

Then, based on the collected papers or materials where the designers explain their design intents, considerations, and procedures, we coded the 61 design projects in three dimensions: (i) design fields, (ii) design tasks, and (iii) design methods.
Specifically, (i) concerns \textit{where} a design problem is formed and originated, (ii) is about \textit{what} goals a design intends to achieve, and (iii) concerns \textit{how} to perform a design. For (iii), we followed the design analysis of narrative visualization~\cite{segel2010narrative} and coded both genres and techniques. \x{Specifically, \textit{genre} refers to a high-level category of design characterized by a particular style or form, while technique refers to low-level methods used to create that design.}
Similar to the coding methodology in ~\cref{sec:argument}, two authors first independently went through the design projects and generated codes and themes regarding the three dimensions. 
\x{To ensure appropriate wording of the techniques, we familiarized ourselves with conventional concepts in design literature. For example, \textit{anthropomorphism} has been used by visualization researchers~\cite{boy2017showing,liem2020structure,ivanov2019walk} to describe the technique of showing concrete human behind data.
After this round of coding, we achieved consensus on 11 (61\%) techniques, such as \textit{color}, \textit{typeface}, and \textit{anthropomorphism}. 
For the remaining techniques, we discussed and refined their wording until achieving 100\% consensus. 
For example, one coder initially used the conventional concept, \textit{animation}, to describe the technique that utilizes motion to convey emotion. However, given that not all the design projects in our corpus are displayed on the screen, \textit{animation} is not suitable to describe the kinetic effects designed for genres such as installations. To enhance the generalizability of this design technique, we finally refined its wording as \textit{kinetic movement}.}

\subsection{Design Space}

After the coding process, we derived the design space in \cref{tab:space}. The table contains 61 rows, each representing one affective visualization design project. The columns represent the three dimensions mentioned above and are divided into several sub-columns. The table uses categorical hues to represent different genres. For other cells, the table uses a check mark colored in grey to indicate the presence of a particular factor or class. Examples of these design projects can be seen in \cref{fig:projects}.

\subsubsection{Where: Design Fields}


\begin{figure*}[t!]
 \centering
 \includegraphics[width=\textwidth]{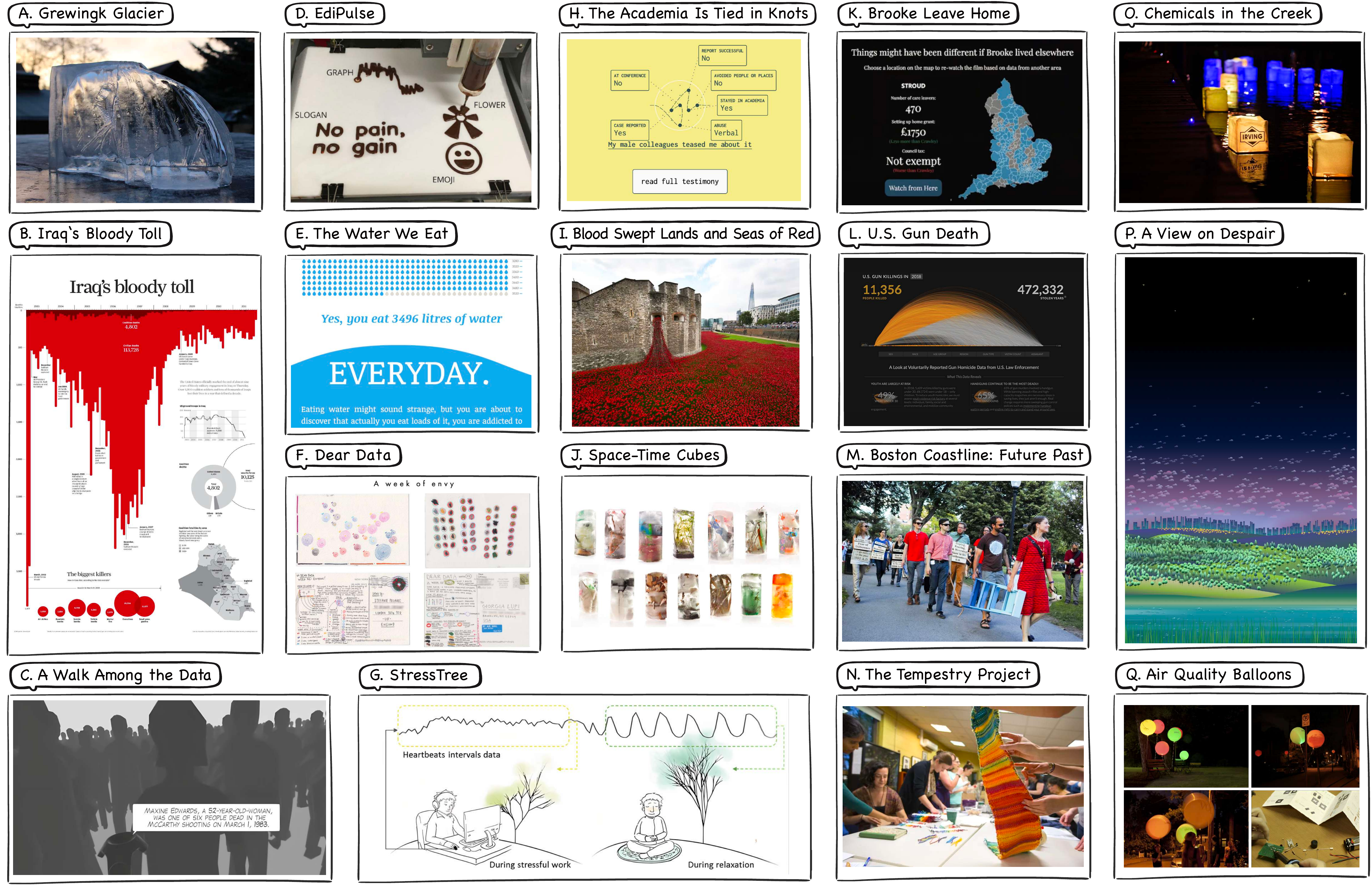}
 \caption{Examples of affective visualization design projects.
 }
 \label{fig:projects}
 \vspace{-2em}
\end{figure*}

Design fields concern where a design need originated, thereby indicating the problem that the design seeks to address and suggesting the scenarios for its application.
Among the 61 design projects, 16 (26\%) projects deal with issues about \textit{environmental sciences \& ecology}. For example, \cref{fig:projects} A, M, N were motivated by global warming. \cref{fig:projects} O and Q address the issue of pollution and \cref{fig:projects} E presents the facts of water waste.
15 (25\%) projects address \textit{social issues}, such as public security, welfare, injustice, or inequality.
For example, \cref{fig:projects} C and L visualize the gun problem in the US.
\cref{fig:projects} B and I visualize the deaths in wars. \cref{fig:projects} H reveals the problem of sexual harassment, and \cref{fig:projects} K focuses on the caring of disadvantaged groups.
11 (18\%) projects are about \textit{health \& well-being}, such as illness and mental stress. 
As an example, \cref{fig:projects} P visualizes suicide data to showcase the problem of depression.
Other fields identified include \textit{news \& media} (7\%), \textit{daily life} (5\%), \textit{business} (3\%), \textit{urban studies} (3\%), and \textit{history} (3\%).
Four projects (7\%) address mixed topics and two projects (3\%) have not specified which fields their design problems originated from.

\subsubsection{What: Design Tasks}

We identified 10 different tasks of affective visualization design. 

\textbf{Inform} (N = 35).
This task concerns communicating certain information to the audience successfully (\eg making people aware of or understand a fact) with the help of evoking emotion.
For example, Grewingk Glacier (\cref{fig:projects} A) is a data-driven installation that visualizes the shrinking speed of the glacier in Alaska. As time passes, the architecture of this installation collapses more quickly, mimicking the fast recession of the glacier while creating a strong emotional impact. 
Through the design, the designer aims to "bridge the gap between reason and emotion" and "synthesize information into knowledge through an intently subjective human experience"~\cite{glacier}.

\textbf{Engage} (N = 27).
This task is about motivating people to consume and spend time with visualization (\eg attracting attention, sparking interest, enticing likes or sharing) through eliciting emotion. For example, Iraq‘s bloody toll (\cref{fig:projects} B) is an infographic that visualizes the number of soldiers who died in the Iraq war. It turns a red bar chart upside down as if blood is dripping down. As explained by the designer, the design choice was made deliberately to create a striking visual and draw readers into the graphic~\cite{iraq}. 

\textbf{Experiment} (N = 14).
This task focuses on realizing the needs of designers themselves to test new design approaches, exercise novel ideas, or stretch the boundaries of affective design. For example, given the new possibilities posed by VR technology, Ivanov~\etal~\cite{ivanov2019walk} created a data story about gun shootings (\cref{fig:projects} C) to explore the elicitation of emotional responses through an immersive environment.

\textbf{Provoke} (N = 12).
This task guides people to think, reflect, contemplate, and inspire new thoughts through emotion.
For example, EdiPulse~\cite{floyd2021eat} (\cref{fig:projects} D) is an edible visualization designed to "provoke critical discourse" about physical activities. The designers translated users' heart rate data into 3D-printed slogans, emojis, and flower-like graphs using chocolate to create a playful and fun experience.

\textbf{Advocate} (N = 11).
This task recommends ideas, values, or proposals to viewers and regards eliciting emotion as a means to persuade participation and action.
For example, The Water We Eat (\cref{fig:projects} E) is an interactive website that advocates action in saving water. This web page visualizes the huge amount of water consumption every day and urges viewers strongly to change their living habits~\cite{water}.

\textbf{Socialize} (N = 11).
This task views emotion as an essential factor for social activities such as sharing, gift-giving, and opinion exchange.
For example, in the Dear Data project~\cite{lupi2016dear}, two designers drew visualizations on postcards based on their personal data collected every week and sent the postcards to each other (\eg \cref{fig:projects} F are postcards drawn to show "A week of envy"). The designers used these visualizations to collaborate and make friends~\cite{deardata}.

\textbf{Heal} (N = 5).
This task intends to increase the well-being of people through the elicitation of positive emotions and the reduction of stress or pain.
For example, StressTree (\cref{fig:projects} G) maps people's stress levels to the colors of a tree metaphorically, thus guiding them to regulate their breathing to make the tree look more healthy. This visualization was designed to help people relax and reduce negative thoughts~\cite{yu2017stresstree}.

\textbf{Empower} (N = 4).
This task focuses on making people (especially vulnerable populations) stronger and more confident in expressing their feelings, controlling their life, and claiming their rights. 
For instance, The Academia is Tied in Knots (\cref{fig:projects} H) surveyed women who have experienced sexual harassment and transformed the data into an interactive interface (the "knots" can be unfolded to show their testimonies) to help bring the sensitive and uneasy stories to the foreground~\cite{elli2020tied}.

\textbf{Commemorate} (N = 3).
This task makes people honor and remember someone or something. 
For example, Blood Swept Lands and Seas of Red~\cite{blood} (\cref{fig:projects} I) installed 888,246 ceramic poppies (each representing a dead soldier) engulfing the Tower of London during the 100th anniversary of Britain’s participation in the World War I. Its designer said, "we couldn’t think of a more appropriate and fitting tribute to the remembrance of those that fell during the World Wars...the Tower of London plays the perfect backdrop to this humbling tribute."~\cite{blood2}.

\textbf{Archive} (N = 2).
This task is about the documentation of personally meaningful experiences and feelings. For example, to archive her walking experience in a park, Gardener~\cite{gardener2017interdisciplinary} captured data (\eg colors, physical objects such as leaves and petals) that triggered her sensory experiences at different locations along the walking route, and then visualized the data as space-time cubes (\cref{fig:projects} J) to facilitate subjective interpretation and recollection.

\subsubsection{How: Design Methods}

This section reports the methods of affective visualization design, including high-level genres and low-level techniques.

\begin{figure}[h]
 \centering
 \vspace{-0.5em}
 \includegraphics[width=\columnwidth]{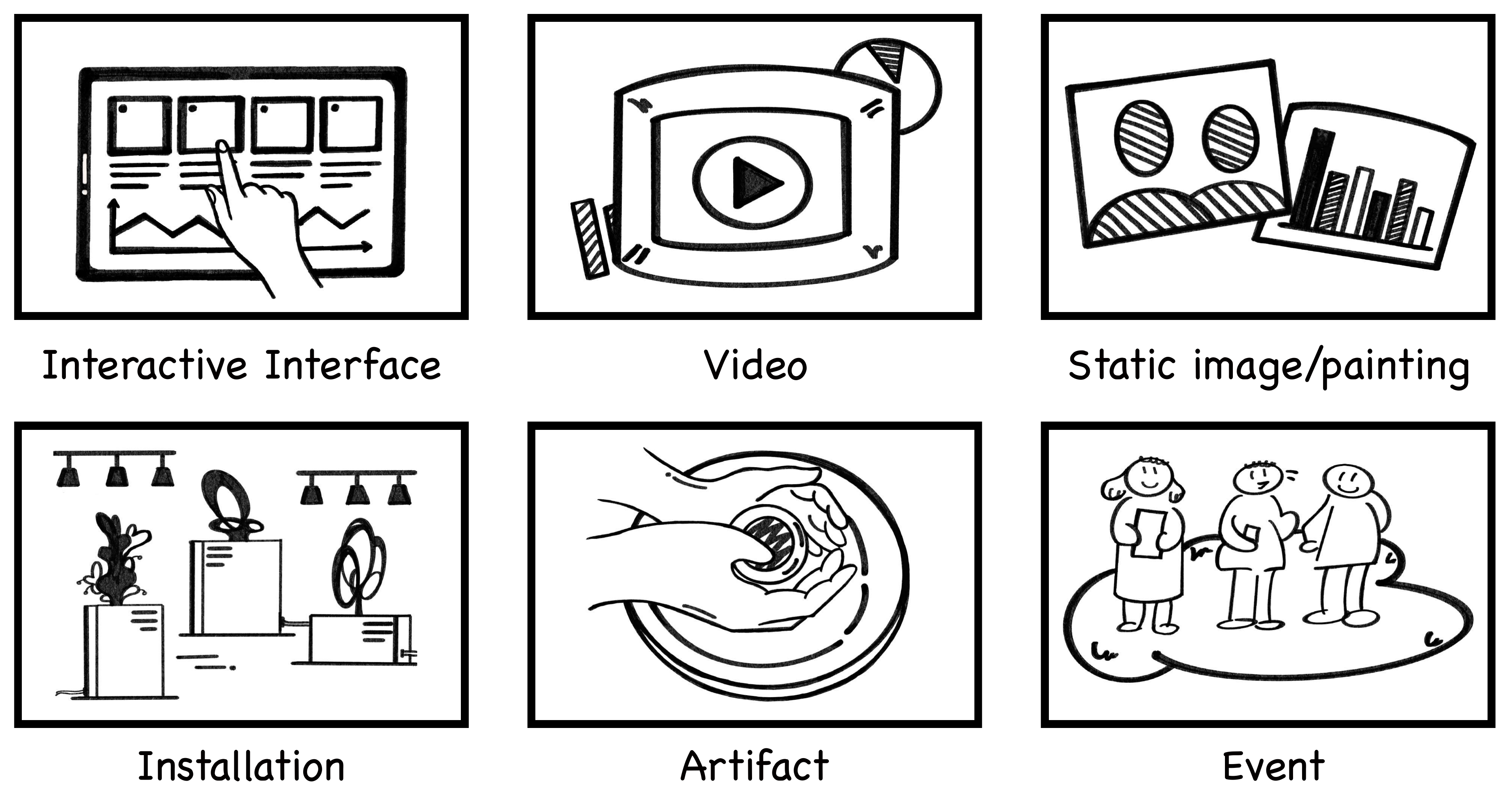}
  \vspace{-2em}
 \caption{Six genres of affective visualization design.}
 \label{fig:genres}
 \vspace{-0.5em}
\end{figure}

\textbf{Genres.}
As illustrated in \cref{fig:genres}, we identified a total of six genres of affective visualization design.
The first one is the \textit{interactive interface}, where users can interact with (\eg click, scroll, rotate) data visualization on web pages or visual systems.
For example, \cref{fig:projects} C, E, H, L all belong to this genre. 
The second one is the \textit{video}, which presents data visualization dynamically as a series of continuous frames. For example, \cref{fig:projects} K is a data film produced to elicit empathy for young adults who leave the care system in England. 
The third genre is the \textit{static image/painting}, such as infographics drawn in computer software and graphs drawn by hands (\eg \cref{fig:projects} B, F).
The fourth genre, \textit{installation}, refers to the data-driven display installed at places such as museums, streets, and galleries. For example, \cref{fig:projects} A, I, Q were all installed outdoors and can be encountered by any passers-by.
The fifth genre is called \textit{artifact}, which denotes the physical object made by hands or devices (\eg a 3D printer). Compared to installations, artifacts are often small objects that are casually used or placed, such as the data-driven scarf in \cref{fig:projects} N and the ice cubes in \cref{fig:projects} J.
The last genre is called \textit{event}, namely an activity intentionally designed for people to participate in and experience data. For example, \cref{fig:projects} M is a “walking data visualization” which asked the participants to walk along the future coastline of a city. Many participants experienced a sense of urgency and concern because the streets they walked on will be flooded by sea water in the future.
Overall, compared to the design genres identified in the previous decade~\cite{segel2010narrative}, the genres of affective visualization design have significantly expanded and diversified.

\begin{table*}[t!]
 \centering
 \includegraphics[width=\textwidth]{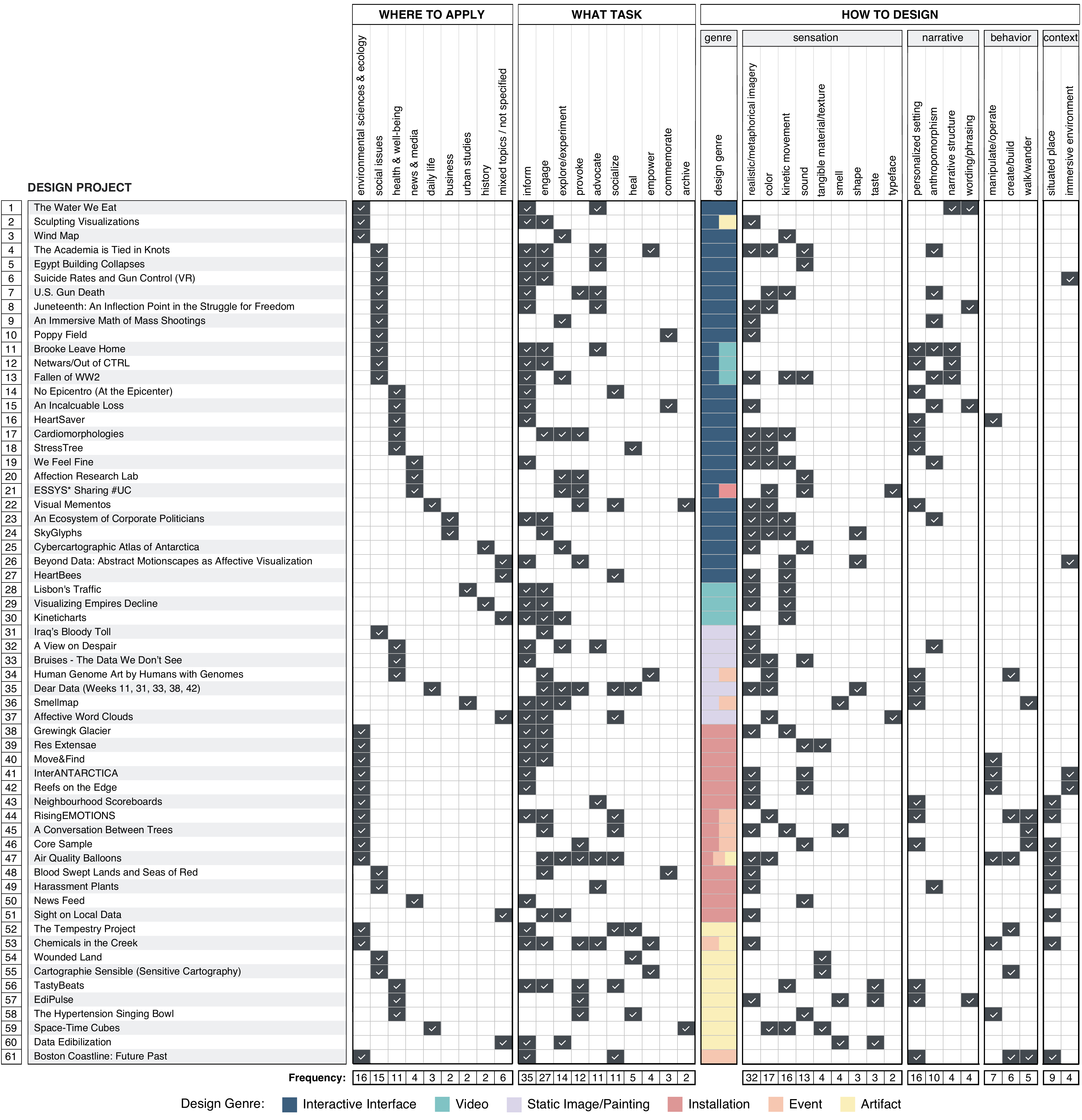}
 \vspace{-2em}
 \caption{Analysis of 61 affective visualization design projects. \x{The splits in the "design genre" column indicate that the corresponding design projects are a combination of two or more different genres.} More detailed information can be browsed at \url{https://affectivevis.github.io/}.}
 \label{tab:space}
 \vspace{-1.5em}
\end{table*}

\textbf{Techniques.}
We identified a total of 18 low-level design techniques used by the 61 affective visualization design projects to influence emotion. We categorized these techniques into four strategies (S1-S4).

S1: Sensation.
Appealing to various senses is ranked as the most applied strategy for affective visualization design. 
In terms of specific design techniques, the use of \textit{realistic/metaphorical imagery} such as photographs, figurative data glyphs, and visual metaphors was mentioned extensively by designers. 
For instance, the installation in \cref{fig:projects} A uses the imagery of ice to evoke viewers' intuitive perception of glaciers. \cref{fig:projects} B presents a bar chart metaphorically to create an imagery of blood. \cref{fig:projects} P uses figurative symbols to represent different types of suicide and deliberately composes them into a peaceful landscape. \cref{fig:projects} I utilizes the symbolic meaning of poppies to mourn for deaths.
Apart from this technique, the use of \textit{color}, \textit{\x{kinetic movement}}, \textit{sound}, \textit{shape}, and \textit{typeface} has also been mentioned by designers as approaches to influencing emotion.
A few projects have also leveraged the emotional quality of \textit{smell} and \textit{taste}.

S2: Narrative.
The second most applied strategy for affective visualization design is narrative, namely using storytelling methods to communicate data affectively.
The most mentioned technique in this category is \textit{\x{personalized setting}}. Given that people generally show stronger emotion to things that relate to themselves, designers have explored various means of improving the sense of proximity. For example, \cref{fig:projects} K lets viewers select their own location and then shows them a customized data film based on the chosen location.
The second most identified technique in this category is \textit{anthropomorphism}, namely showing the real people behind data~\cite{morais2022exploring}. For example, \cref{fig:projects} H uses one glyph to show one testimony from a victim of sexual harassment, and \cref{fig:projects} C allows viewers to "meet" the casualties of gun shootings and read their identities, thus transforming abstract numbers into people with flesh and blood.
In addition, designers have adopted techniques such as \textit{narrative structure} and \textit{wording/phrasing} to manipulate the flow of data narrative and augment emotion through textual expressions.

S3: Behavior.
Behavioral design is a design method that seeks to shape or influence human behavior.
During our analysis, we identified three main techniques of behavior design. 
The first technique asks people to \textit{manipulate/operate} a data visualization. For example, the designers of \cref{fig:projects} O prepared a set of river lanterns that encoded the data of river pollution in a community and invited the residents to release the lanterns into the river, thus engaging the residents emotionally with the data.
The second technique asks people to \textit{create/build} a data visualization by themselves, such as the knitting of data-driven scarves in \cref{fig:projects} N.
The third technique makes people experience data during a \textit{walk/wander}. For example, 
\cref{fig:projects} M asked people to walk along the route of the future coastline to subjectively experience climate change, and \cref{fig:projects} Q asked people to step into parks, streets, and neighborhoods to monitor air quality in an embodied manner.

S4: Context.
Context means the surrounding environment when people view or interact with a data visualization. For example, a viewer may read a data visualization from home or on the road, which can distance him/her from the content being portrayed in the visualization.
To cope with such disconnection, several design projects put a visualization at a \textit{situated place} (\ie somewhere related to the content being presented in a data visualization design).
For example, apart from creating visualizations about river pollution, \cref{fig:projects} O also brought local residents to the polluted river to heighten the emotional impact. \cref{fig:projects} I chose the Tower of London as the place to install the poppies because it is a representation of national honor and is related to the theme of the data visualization, namely sacrificing for the country.
We also found three projects that used VR technology or large screen projection technology to create an \textit{immersive environment}, such as \cref{fig:projects} C.

\subsubsection{Design Space Observations}

By observing how the cells distribute in \cref{tab:space}, we summarize implications from three main aspects. 

\textbf{Task distribution.}
The most mentioned tasks of affective visualization design are \textit{inform} (mentioned by 57\% of all projects) and \textit{engage} (44\%), followed by \textit{experiment} (23\%),  \textit{provoke} (20\%), \textit{advocate} (18\%), \textit{socialize} (18\%), \textit{heal} (8\%), \textit{empower} (7\%), \textit{commemorate} (5\%), and \textit{archive} (3\%).
Most projects claimed hybrid tasks, and the average number of tasks undertaken per design project is 2.03 (SD = 1.06). 
Two projects claimed as many as five tasks.
For example, Dear Data~\cite{lupi2016dear} project (\cref{tab:space} \#35) was initiated by the motivation to make friends (\textit{socialize}) and \textit{experiment} how to visualize personal data through hand-drawing postcards. During the drawing process, the two designers also had more reflections on their own lives (\textit{provoke}) and used the drawing as a therapy to face themselves honestly (\textit{heal}). Meanwhile, they also published the postcards to \textit{engage} more people with data visualization~\cite{deardata}. 
We also calculated how these design tasks interact with different design fields. For example, in the field of \textit{environmental sciences \& ecology}, 69\% projects reported the \textit{inform} task, 50\% reported the \textit{engage} task, and 31\% reported the \textit{socialize} task.
In the field of \textit{social issues}, 53
\% projects reported the \textit{inform} task, 47\% reported the \textit{engage} task, and 40\% reported the \textit{advocate} task. 
When designing for \textit{daily life}, 67\% projects intended to \textit{archive}, \textit{provoke}, and \textit{socialize}.
In general, although \textit{inform} and \textit{engage} are the dominant tasks in most cases, different fields still show varied priorities in design tasks.

\textbf{Genre distribution.} 
The most popular affective visualization design genre in our corpus is the \textit{interactive interface} (adopted by 44\% of all projects), followed by \textit{installation} (25\%), \textit{artifact} (18\%), \textit{event} (13\%), \textit{static image/painting} (11\%), and \textit{video} (10\%).
Although most design projects belong to a single genre, we identified 12 projects that belong to hybrid genres.
For example, three projects (\cref{tab:space} \#11-13) published two versions of their design, including an \textit{interactive interface} and a \textit{video}.
Three projects (\cref{tab:space} \#44-46) created a data-driven \textit{installation} while organizing an outdoor \textit{event}.
Project \#47 is the only design work that spans three genres. Its designers first set a data-driven \textit{installation} (\ie a cluster of balloon sensors that monitor air quality) at public sites, and then distributed the balloons to volunteers and asked them to wander in the city and play with the balloons (\textit{event}). In addition, they also organized a workshop for people to create the balloons by themselves (\textit{artifact}).
As for the interaction between genres and design fields, \textit{interactive interface} is the primary genre within most fields, such as \textit{news \& media} (adopted by 75\% projects within this field), \textit{social issues} (67\%), and \textit{health \& well-being} (45\%). However, for \textit{environmental sciences \& ecology}, compared to \textit{interactive interface} (19\%), the dominant genre is \textit{installation} (63\%). 
One reason may be that environmental issues are closely related to the physical world, so many designers decided to invite people to experience data in outdoor space rather than just showing data on screens.


\textbf{Technique distribution.} 
Overall, sensation is the most applied strategy to elicit emotion, and among all the techniques within this strategy, the most adopted techniques are \textit{realistic/metaphorical imagery} (adopted by 52\% projects), \textit{color} (28\%), \textit{\x{kinetic movement}} (26\%), and \textit{sound} (21\%). 
The primary sense being designed is the visual sense, followed by the auditory sense.
The strategy of narrative has also been practiced by many projects, with \textit{\x{personalized setting}} (26\%) and \textit{anthropomorphism} (16\%) being the most frequently applied techniques.
The two techniques both serve to enhance the individuality of data, either by making it relevant to viewers themselves or by showing other people behind data.
It has also been observed that some techniques are preferred by specific genres. For example, genres such as \textit{interactive interface}, \textit{video}, and \textit{static image/painting} often evoke emotion through sensation and narrative.
By contrast, genres such as \textit{event} and \textit{artifact} tend to appeal more to sensation and behavior. 
This distinction may be attributed to the nature of these genres, as the former three genres are more digital, whereas the latter two genres are more physical.
However, we did not find an absolute boundary between these design genres in terms of using design techniques.
For example, digital genres can use VR technology to create a sense of physical context (\eg \cref{tab:space} \#6) or employ interactive features to manipulate user behavior online (\eg ~\cref{tab:space} \#16). Likewise, physical artifacts can also integrate digital elements (\eg sensors, 3D printing) to dynamically generate personal narratives (\eg \cref{tab:space} \#56, 57). 
These design practices suggest that the fusion of online and offline experiences and the exploration of cross-media design are taking place in affective visualization design.

\section{Discussion and Future Work}
\label{sec:discussion}

Below we discuss our limitations and future research opportunities. 

\subsection{Limitations}
In the corpus collection stage, we used the snowball search method to collect as much qualified literature as possible. \x{However, the corpus may not be exhaustive as the snowball searching relies on the references cited in the initial set of papers, potentially overlooking less-cited sources that are not referenced in the initial set.}
Secondly, the 61 design projects used for characterizing the design space of affective visualization design 
\x{are either research papers themselves or in-the-wild projects cited by research papers.}
Therefore, although being typical examples of affective visualization design in the eyes of researchers, they are by no means exhaustive nor representative of the full landscape of affective visualization design practice. 
\x{We believe that more case studies and interviews with practitioners can be conducted to help identify and analyze the affective visualization design practice in the wild.}
Also, alternative analysis perspectives or taxonomies can be proposed to further advance our understanding of such design practice.

\subsection{Opportunities For Future Research}
\x{We identify nine specific research opportunities (marked as O1-O9) categorized into three main avenues.}

\textbf{Scientific evaluation of affective design techniques.}
The design space in \cref{sec:space} indicates a set of design techniques that designers thought can influence emotion. However, as shown in prior work~\cite{lan2022negative}, sometimes there will be a gap between designers' expectations and people's real experiences. 
Among the identified techniques, although some have been validated by controlled user studies as being effective in influencing emotion, such as color and animation~\cite{bartram2017affective,lan2021kineticharts,lan2022chart}, a considerate number of techniques still lack evaluation. \x{Therefore, more future work needs to be done to \ul{(O1) assess the relationship between visualization design techniques and emotion.}}
\x{Second, controversy still exists in terms of the effectiveness of certain techniques.} For example, anthropomorphism was found insignificant in augmenting empathy by Boy~\etal~\cite{boy2017showing}, while it showed a small effect in another larger-scale study~\cite{morais2021can}. In a study about map design, people got significantly more emotionally involved with locations close to themselves~\cite{gilmartin1991effects}, but in another study, \x{such a personalized narrative} showed no significant effect~\cite{concannon2020brooke}. \x{To ensure the reliability of research findings, future studies can \ul{(O2) replicate existing results in diverse settings and cross-validate conflicting outcomes.}}
Third, as more and more state-of-the-art technology has been incorporated into affective visualization design, \x{the design space of affective visualization design will go on to expand. It is important to study \ul{(O3) how novel techniques such as AR/VR/MR, 3D printing, and multi-modal interaction (\eg sound, smell) influence emotion and how they interact with other design elements}}.
\x{Fourth, more efforts can be put into enhancing the rigor of research methods. Researchers can \ul{(O4) conduct studies with more diverse stimuli and larger sample sizes, and try alternative measures (\eg neuro-physiological measures} to understand emotion more in-depth.}



\x{\textbf{Embracing more application scenarios and users.}}
Through this work, we are excited to see affective visualization design being applied to various domains and being used as a novel approach to solving domain problems and serving realistic needs. We see journalists, non-profits, and feminists use data visualization to raise awareness of social injustice. We see environmentalists and geographers use data visualization to call for green lifestyles. We see artists create data visualization to touch the tender part of human hearts. We see scientists transform complex, cold data into vivid, affective representations. \x{In the future, we believe that \ul{(O5) the idea of affective visualization design can be introduced to a wider range of disciplines and industries, leading to more interesting applications.}} 
Another important lesson we have learned from this work is that, rather than seeing people as passive information receivers, many design projects have treated people as active agents or collaborators. Such a paradigm has been framed by some researchers and designers as \textit{participatory action research}, \textit{citizen science}, or \textit{grassroots movements}~\cite{kuznetsov2011red,aragon2021risingemotions,perovich2020chemicals}.
Under this paradigm, designers in fact relinquish some design rights to people and allow them to participate in or even dominate the process of data collection and visualization. 
In the future, we may further \x{\ul{(O6) explore the approaches to stimulating people's sense of agency in data visualization design and bringing data democratization to a higher level}}. 
\x{Relating to this issue, \ul{(O7) more authoring tools can be developed to empower ordinary people to create or customize affective visualization design.} Currently, this type of tool is surprisingly rare~\cite{xie2023creating,romat2020dear,chen2023does}.}



\x{\textbf{Evoking emotion appropriately and ethically.}
Although this work focuses on presenting the arguments and design projects that support evoking emotion with data visualization design, we do not claim that eliciting emotion is always good or that objectivity and neutrality should be tossed away.
Emotion is a double-edged sword, and an important future research direction is to \ul{(O8) examine the pitfalls or side effects of eliciting emotion with data visualization.} For example, emotion may make people impulsive and prejudiced. Psychologist Bloom~\cite{bloom} argued for putting empathy aside in public decision-making because "empathy is biased; we are more prone to feel empathy for attractive people and for those who look like us or share our ethnic or national background...and empathy is narrow; it connects us to particular individuals, real or imagined, but is insensitive to numerical differences and statistical data." Another viewpoint is embedding a pre-defined emotion in a visualization may hinder viewers' freedom to interpret the design emotionally on their own terms. For example, cartographer Kent~\cite{kent2013dry} thought that “while the emotional association with a specific place would perhaps be affected by the amount of detail apparent in its portrayal...the absence of detail inherent to cartographic symbolization allows a free play of the imagination necessary for the development of emotions associated with that sense of place”.
In addition, the power of emotion may be exploited by people with malevolent intents. For example, social science has a long history of examining how politicians and businessmen incite emotion to gain profit or deceive people with low literacy, as well as how emotional contagion aids the spread of violent or destructive behavior~\cite{le1897crowd}. Therefore, it is crucial to \ul{(O9) study the ethical issues surrounding affective visualization design and develop guidelines that promote the responsible use of such design.}}

\section{Conclusion}

In this work, we \x{have identified three distinct research threads that study visualization and emotion and defined affective visualization design as data visualizations designed to communicate and influence emotion}. We have reviewed 109 papers concerning affective visualization design to characterize this growing field and outline the necessity of considering emotion in visualization design. We have also analyzed 61 affective visualization design projects to explore the design space of such practices, including where to apply the design, what tasks the design can undertake, and how to perform the design.

Affective visualization design is an ongoing research field that is full of creativity and innovation. It responds to various emerging needs in the real world and also deepens our understanding of data visualization, such as what constitutes a good data visualization, how a visualization should be evaluated, and what visualization can do for individuals and society.
\x{}
We believe that more investigation and discussion of affective visualization design will help the visualization community embrace more ideas, values, application scenarios, and wider users. 


\acknowledgments{
This work was supported by NSFC 62072338 and NSF Shanghai 20ZR1461500. We would like to thank all the reviewers for their valuable feedback.}

\bibliographystyle{abbrv-doi-hyperref}

\bibliography{template}

\begin{thebibliography}{10}

\bibitem{apa}
Affect.
\newblock \url{https://dictionary.apa.org/affect}.
\newblock \x{Last accessed: June 25, 2023}.

\bibitem{area}
Area model for vis.
\newblock
  \url{https://ieeevis.org/year/2022/info/call-participation/area-model}.
\newblock \x{Last accessed: June 25, 2023}.

\bibitem{wos}
Web of science research areas.
\newblock
  \url{https://images.webofknowledge.com/images/help/WOS/hp_research_areas_easca.html}.
\newblock \x{Last accessed: June 25, 2023}.

\bibitem{aitken2011affective}
S.~Aitken and J.~Craine.
\newblock Affective geovisualisations.
\newblock In {\em The Map Reader: Theories of Mapping Practice and Cartographic
  Representation}, pp. 278--280. Wiley Online Library, 2011.

\bibitem{ajani2021declutter}
K.~Ajani, E.~Lee, C.~Xiong, C.~N. Knaflic, W.~Kemper, and S.~Franconeri.
\newblock Declutter and focus: Empirically evaluating design guidelines for
  effective data communication.
\newblock {\em IEEE Transactions on Visualization and Computer Graphics},
  28(10):3351--3364, 2021.

\bibitem{alamalhodaei2020humanizing}
A.~Alamalhodaei, A.~P. Alberda, and A.~Feigenbaum.
\newblock Humanizing data through ‘data comics’: An introduction to graphic
  medicine and graphic social science.
\newblock In {\em Data Visualization in Society}, pp. 347--365. Amsterdam
  University Press, 2020.

\bibitem{anderson2021affective}
C.~L. Anderson and A.~C. Robinson.
\newblock Affective congruence in visualization design: Influences on reading
  categorical maps.
\newblock {\em IEEE Transactions on Visualization and Computer Graphics},
  28(8):2867--2878, 2021.

\bibitem{aragon2021risingemotions}
C.~Arag{\'o}n, M.~Jasim, and N.~Mahyar.
\newblock \x{{RisingEMOTIONS}}: Bridging art and technology to visualize
  public’s emotions about climate change.
\newblock In {\em Creativity and Cognition}, pp. 1--10. ACM, 2021.

\bibitem{aseniero2022skyglyphs}
B.~A. Aseniero, S.~Carpendale, G.~Fitzmaurice, and J.~Matejka.
\newblock Skyglyphs: Reflections on the design of a delightful visualization.
\newblock In {\em IEEE VIS Arts Program}, pp. 105--120. IEEE, 2022.

\bibitem{bartram2017affective}
L.~Bartram, A.~Patra, and M.~Stone.
\newblock Affective color in visualization.
\newblock In {\em Proceedings of the CHI Conference on Human Factors in
  Computing Systems}, pp. 1364--1374. ACM, 2017.

\bibitem{beale2008role}
R.~Beale and C.~Peter.
\newblock The role of affect and emotion in hci.
\newblock In {\em Affect and Emotion in Human-Computer Interaction}, pp. 1--11.
  Springer, 2008.

\bibitem{bloom}
P.~Bloom.
\newblock Against empathy.
\newblock \url{https://www.bostonreview.net/forum/paul-bloom-against-empathy/},
  2014.
\newblock \x{Last accessed: June 25, 2023}.

\bibitem{boy2017showing}
J.~Boy, A.~V. Pandey, J.~Emerson, M.~Satterthwaite, O.~Nov, and E.~Bertini.
\newblock Showing people behind data: Does anthropomorphizing visualizations
  elicit more empathy for human rights data?
\newblock In {\em Proceedings of the CHI Conference on Human Factors in
  Computing Systems}, pp. 5462--5474. ACM, 2017.

\bibitem{braun2012thematic}
V.~Braun and V.~Clarke.
\newblock {\em Thematic analysis.}
\newblock American Psychological Association, 2012.

\bibitem{campbell2018feeling}
S.~Campbell.
\newblock Feeling numbers: the rhetoric of pathos in visualization.
\newblock Master's thesis, 2018.
\newblock Northeastern University.

\bibitem{carpendale2017subjectivity}
S.~Carpendale, A.~Thudt, C.~Perin, and W.~Willett.
\newblock Subjectivity in personal storytelling with visualization.
\newblock {\em Information Design Journal}, 23(1):48--64, 2017.

\bibitem{cartwright2008developing}
W.~Cartwright, A.~Miles, B.~Morris, L.~Vaughan, and J.~Yuille.
\newblock Developing concepts for an affective atlas.
\newblock In {\em Geospatial Vision}, pp. 219--234. Springer, 2008.

\bibitem{chen2023does}
Q.~Chen, S.~Cao, J.~Wang, and N.~Cao.
\newblock How does automation shape the process of narrative visualization: A
  survey of tools.
\newblock {\em IEEE Transactions on Visualization and Computer Graphics}, 2023.

\bibitem{claes2017public}
S.~Claes and A.~Vande~Moere.
\newblock What public visualization can learn from street art.
\newblock {\em Leonardo}, 50(1):90--91, 2017.

\bibitem{concannon2020brooke}
S.~Concannon, N.~Rajan, P.~Shah, D.~Smith, M.~Ursu, and J.~Hook.
\newblock Brooke leave home: Designing a personalized film to support public
  engagement with open data.
\newblock In {\em Proceedings of the CHI Conference on Human Factors in
  Computing Systems}, pp. 1--14. ACM, 2020.

\bibitem{craine2011emotional}
J.~Craine and S.~C. Aitken.
\newblock The emotional life of maps and other visual geographies.
\newblock In {\em Rethinking Maps}, pp. 167--184. Routledge, 2011.

\bibitem{blood}
P.~Cummins.
\newblock Blood swept lands and seas of red.
\newblock \url{https://www.paulcumminsceramics.com/blood-swept/}, 2014.
\newblock \x{Last accessed: June 25, 2023}.

\bibitem{deardata}
DataStories.
\newblock \x{"{Dear} {Data}" with {Giorgia} {Lupi} and {Stefanie} {Posavec}}.
\newblock
  \url{https://datastori.es/dear-data-with-giorgia-lupi-and-stefanie-posavec-ds64/},
  2015.
\newblock \x{Last accessed: June 25, 2023}.

\bibitem{d2020data}
C.~D'ignazio and L.~F. Klein.
\newblock {\em Data feminism}.
\newblock MIT press, 2020.

\bibitem{dragga2001cruel}
S.~Dragga and D.~Voss.
\newblock Cruel pies: The inhumanity of technical illustrations.
\newblock {\em Technical Communication}, 48(3):265--274, 2001.

\bibitem{elli2020tied}
T.~Elli, A.~Bradley, C.~Collins, U.~Hinrichs, Z.~Hills, and K.~Kelsky.
\newblock Tied in knots: A case study on anthropographic data visualization
  about sexual harassment in the academy.
\newblock In {\em IEEE VIS Arts Program}, pp. 29--44. IEEE, 2020.

\bibitem{floyd2021eat}
F.~'Floyd'Mueller, T.~Dwyer, S.~Goodwin, K.~Marriott, J.~Deng, H.~D.~Phan,
  J.~Lin, K.-T. Chen, Y.~Wang, and R.~Ashok~Khot.
\newblock Data as delight: Eating data.
\newblock In {\em Proceedings of the CHI Conference on Human Factors in
  Computing Systems}, pp. 1--14. ACM, 2021.

\bibitem{gardener2017interdisciplinary}
J.~Gardener, W.~Cartwright, and L.~Duxbury.
\newblock An interdisciplinary approach to mapping through scientific
  cartography, design and artistic expression.
\newblock In {\em Proceedings of the International Cartographic Association},
  pp. 1--6, 2017.

\bibitem{gilmartin1991effects}
P.~Gilmartin and R.~Lloyd.
\newblock The effects of map projections and map distance on emotional
  involvement with places.
\newblock {\em The Cartographic Journal}, 28(2):145--151, 1991.

\bibitem{gough2014affective}
P.~Gough, C.~de~Berigny~Wall, and T.~Bednarz.
\newblock Affective and effective visualisation: Communicating science to
  non-expert users.
\newblock In {\em IEEE Pacific Visualization Symposium}, pp. 335--339. IEEE,
  2014.

\bibitem{harrison2013influencing}
L.~Harrison, D.~Skau, S.~Franconeri, A.~Lu, and R.~Chang.
\newblock Influencing visual judgment through affective priming.
\newblock In {\em Proceedings of the CHI Conference on Human Factors in
  Computing Systems}, pp. 2949--2958. ACM, 2013.

\bibitem{water}
InfoDesignLab.
\newblock The water we eat.
\newblock \url{http://thewaterweeat.com/}, 2012.
\newblock \x{Last accessed: June 25, 2023}.

\bibitem{iturrioz2011artistic}
T.~Iturrioz and M.~Wachowicz.
\newblock An artistic perspective for affective cartography.
\newblock In {\em Mapping Different Geographies}, pp. 74--92. Springer, 2011.

\bibitem{ivanov2019walk}
A.~Ivanov, K.~Danyluk, C.~Jacob, and W.~Willett.
\newblock A walk among the data: Exploration and anthropomorphism in immersive
  unit visualizations.
\newblock {\em IEEE Computer Graphics and Applications}, 39(3):19--28, 2019.

\bibitem{johnson2003effective}
D.~Johnson and J.~Wiles.
\newblock Effective affective user interface design in games.
\newblock {\em Ergonomics}, 46(13-14):1332--1345, 2003.

\bibitem{jorgenson1995is}
L.~J{\"o}rgenson, R.~Kriz, B.~Mones-Hattal, B.~Rogowitz, and F.~D. Fracchia.
\newblock Is visualization struggling under the myth of objectivity?
\newblock In {\em IEEE Visualization Conference}, 1995.

\bibitem{kauer2021public}
T.~Kauer, M.~D{\"o}rk, A.~L. Ridley, and B.~Bach.
\newblock The public life of data: Investigating reactions to visualizations on
  reddit.
\newblock In {\em Proceedings of the CHI Conference on Human Factors in
  Computing Systems}, pp. 1--12. ACM, 2021.

\bibitem{kennedy2018feeling}
H.~Kennedy and R.~L. Hill.
\newblock The feeling of numbers: Emotions in everyday engagements with data
  and their visualisation.
\newblock {\em Sociology}, 52(4):830--848, 2018.

\bibitem{kennedy2016work}
H.~Kennedy, R.~L. Hill, G.~Aiello, and W.~Allen.
\newblock The work that visualisation conventions do.
\newblock {\em Information, Communication \& Society}, 19(6):715--735, 2016.

\bibitem{kent2013dry}
A.~Kent.
\newblock From a dry statement of facts to a thing of beauty: understanding
  aesthetics in the mapping and counter-mapping of place.
\newblock {\em Cartographic Perspectives}, 73:39--60, 2013.

\bibitem{kosara2007visualization}
R.~Kosara.
\newblock Visualization criticism-the missing link between information
  visualization and art.
\newblock In {\em International Conference Information Visualization}, pp.
  631--636. IEEE, 2007.

\bibitem{kostelnick2016re}
C.~Kostelnick.
\newblock The re-emergence of emotional appeals in interactive data
  visualization.
\newblock {\em Technical Communication}, 63(2):116--135, 2016.

\bibitem{kuznetsov2011red}
S.~Kuznetsov, G.~N. Davis, E.~Paulos, M.~D. Gross, and J.~C. Cheung.
\newblock Red balloon, green balloon, sensors in the sky.
\newblock In {\em Proceedings of the International Conference on Ubiquitous
  Computing}, pp. 237--246, 2011.

\bibitem{lan2021kineticharts}
X.~Lan, Y.~Shi, Y.~Wu, X.~Jiao, and N.~Cao.
\newblock Kineticharts: Augmenting affective expressiveness of charts in data
  stories with animation design.
\newblock {\em IEEE Transactions on Visualization and Computer Graphics},
  28(1):933--943, 2022.

\bibitem{lan2021smile}
X.~Lan, Y.~Shi, Y.~Zhang, and N.~Cao.
\newblock Smile or scowl looking at infographic design through the affective
  lens.
\newblock {\em IEEE Transactions on Visualization and Computer Graphics},
  27(6):2796--2807, 2021.

\bibitem{lan2022chart}
X.~Lan, Y.~Wu, Q.~Chen, and N.~Cao.
\newblock The chart excites me! exploring how data visualization design
  influences affective arousal.
\newblock {\em arXiv preprint arXiv:2211.03296}, 2022.

\bibitem{lan2022negative}
X.~Lan, Y.~Wu, Y.~Shi, Q.~Chen, and N.~Cao.
\newblock Negative emotions, positive outcomes? exploring the communication of
  negativity in serious data stories.
\newblock In {\em Proceedings of the CHI Conference on Human Factors in
  Computing Systems}, pp. 1--14. ACM, 2022.

\bibitem{lan2021understanding}
X.~Lan, X.~Xu, and N.~Cao.
\newblock Understanding narrative linearity for telling expressive
  time-oriented stories.
\newblock In {\em Proceedings of the CHI Conference on Human Factors in
  Computing Systems}, pp. 1--13. ACM, 2021.

\bibitem{le1897crowd}
G.~Le~Bon.
\newblock {\em The crowd: A study of the popular mind}.
\newblock T. Fisher Unwin, London, UK, 1897.

\bibitem{lee2020data}
B.~Lee, D.~Brown, B.~Lee, C.~Hurter, S.~Drucker, and T.~Dwyer.
\newblock Data visceralization: Enabling deeper understanding of data using
  virtual reality.
\newblock {\em IEEE Transactions on Visualization and Computer Graphics},
  27(2):1095--1105, 2020.

\bibitem{lee2022affective}
E.~Lee-Robbins and E.~Adar.
\newblock Affective learning objectives for communicative visualizations.
\newblock {\em IEEE Transactions on Visualization and Computer Graphics},
  29(1):1--11, 2022.

\bibitem{lewis2010handbook}
M.~Lewis, J.~M. Haviland-Jones, and L.~F. Barrett.
\newblock {\em Handbook of emotions}.
\newblock Guilford Press, 2010.

\bibitem{liem2020structure}
J.~Liem, C.~Perin, and J.~Wood.
\newblock Structure and empathy in visual data storytelling: Evaluating their
  influence on attitude.
\newblock {\em Computer Graphics Forum}, 39(3):277--289, 2020.

\bibitem{lupi}
G.~Lupi.
\newblock Data humanism, the revolution will be visualized.
\newblock
  \url{http://giorgialupi.com/data-humanism-my-manifesto-for-a-new-data-wold},
  2017.
\newblock \x{Last accessed: June 25, 2023}.

\bibitem{lupi2016dear}
G.~Lupi and S.~Posavec.
\newblock {\em Dear data}.
\newblock Chronicle books, 2016.

\bibitem{mccleary2003beyond}
G.~McCleary.
\newblock Beyond visualization: mapping genocide.
\newblock In {\em Proceedings of the International Cartographic Conference},
  pp. 1827--1834. ICA, 2003.

\bibitem{morais2022exploring}
L.~Morais, N.~Andrade, and D.~Sousa.
\newblock Exploring how visualization design and situatedness evoke compassion
  in the wild.
\newblock {\em Computer Graphics Forum}, 41(3):441--452, 2022.

\bibitem{morais2021can}
L.~Morais, Y.~Jansen, N.~Andrade, and P.~Dragicevic.
\newblock Can anthropographics promote prosociality? a review and large-sample
  study.
\newblock In {\em Proceedings of the CHI Conference on Human Factors in
  Computing Systems}, pp. 1--18. ACM, 2021.

\bibitem{muehlenhaus2012if}
I.~Muehlenhaus.
\newblock If looks could kill: The impact of different rhetorical styles on
  persuasive geocommunication.
\newblock {\em The Cartographic Journal}, 49(4):361--375, 2012.

\bibitem{nietzsche2022gay}
F.~Nietzsche.
\newblock {\em The gay science}.
\newblock DigiCat, 2022.

\bibitem{norman2004emotional}
D.~A. Norman.
\newblock {\em Emotional design: Why we love (or hate) everyday things}.
\newblock Basic Books, 2004.

\bibitem{padilla2022multiple}
L.~Padilla, R.~Fygenson, S.~C. Castro, and E.~Bertini.
\newblock Multiple forecast visualizations (mfvs): Trade-offs in trust and
  performance in multiple covid-19 forecast visualizations.
\newblock {\em IEEE Transactions on Visualization and Computer Graphics},
  29(1):12--22, 2022.

\bibitem{gun2}
Periscopic.
\newblock Revealing the overwhelming magnitude of loss from \x{{U.S.}} gun
  deaths.
\newblock
  \url{https://periscopic.com/\#!/impacts/stolen-years\&impact-area=awareness}.
\newblock \x{Last accessed: June 25, 2023}.

\bibitem{gun}
Periscopic.
\newblock \x{{U.S.}} gun deaths.
\newblock \url{https://guns.periscopic.com/}, 2013.
\newblock \x{Last accessed: June 25, 2023}.

\bibitem{perovich2020chemicals}
L.~J. Perovich, S.~A. Wylie, and R.~Bongiovanni.
\newblock Chemicals in the creek: designing a situated data physicalization of
  open government data with the community.
\newblock {\em IEEE Transactions on Visualization and Computer Graphics},
  27(2):913--923, 2020.

\bibitem{picard2000affective}
R.~W. Picard.
\newblock {\em Affective computing}.
\newblock MIT press, 2000.

\bibitem{pinilla2021affective}
A.~Pinilla, J.~Garcia, W.~Raffe, J.-N. Voigt-Antons, R.~P. Spang, and
  S.~M{\"o}ller.
\newblock Affective visualization in virtual reality: An integrative review.
\newblock {\em Frontiers in Virtual Reality}, 2, 2021.

\bibitem{gun3}
PolicyViz.
\newblock Episode \#31: \x{{Rees} \& {Mushon} on {DataViz}} empathy.
\newblock \url{https://policyviz.com/podcast/rees-mushon-on-dataviz-empathy/},
  2016.
\newblock \x{Last accessed: June 25, 2023}.

\bibitem{porter1996trust}
T.~M. Porter.
\newblock Trust in numbers: the pursuit of objectivity in science \& public
  life.
\newblock In {\em Trust in Numbers}. Princeton University Press, 1996.

\bibitem{qin2020heartbees}
C.~Y. Qin, M.~Constantinides, L.~M. Aiello, and D.~Quercia.
\newblock Heartbees: Visualizing crowd affects.
\newblock In {\em IEEE VIS Arts Program}, pp. 1--8. IEEE, 2020.

\bibitem{rebelo2022essys}
S.~M. Rebelo, M.~Sei{\c{c}}a, P.~Martins, J.~Bicker, and P.~Machado.
\newblock Essys* sharing\# uc: An emotion-driven audiovisual installation.
\newblock In {\em IEEE VIS Arts Program}, pp. 70--79. IEEE, 2022.

\bibitem{romat2020dear}
H.~Romat, N.~Henry~Riche, C.~Hurter, S.~Drucker, F.~Amini, and K.~Hinckley.
\newblock Dear pictograph: Investigating the role of personalization and
  immersion for consuming and enjoying visualizations.
\newblock In {\em Proceedings of the CHI Conference on Human Factors in
  Computing Systems}, pp. 1--13. ACM, 2020.

\bibitem{sallam2022towards}
S.~Sallam, Y.~Sakamoto, J.~Leboe-McGowan, C.~Latulipe, and P.~Irani.
\newblock Towards design guidelines for effective health-related data videos:
  An empirical investigation of affect, personality, and video content.
\newblock In {\em Proceedings of the CHI Conference on Human Factors in
  Computing Systems}, pp. 1--22. ACM, 2022.

\bibitem{samsel2021affective}
F.~Samsel, G.~Abram, S.~Zeller, and D.~Keefe.
\newblock Affective palettes for scientific visualization: Grounding
  environmental data in the natural world.
\newblock In {\em IEEE VIS Arts Program}, pp. 20--34. IEEE, 2021.

\bibitem{iraq}
S.~Scarr.
\newblock Iraq‘s bloody toll.
\newblock
  \url{https://www.scmp.com/infographics/article/1284683/iraqs-bloody-toll},
  2011.
\newblock \x{Last accessed: June 25, 2023}.

\bibitem{glacier}
A.~Segal.
\newblock Grewingk glacier.
\newblock \url{https://www.adriensegal.com/grewingk-glacier}, 2015.
\newblock \x{Last accessed: June 25, 2023}.

\bibitem{segel2010narrative}
E.~Segel and J.~Heer.
\newblock Narrative visualization: Telling stories with data.
\newblock {\em IEEE Transactions on Visualization and Computer Graphics},
  16(6):1139--1148, 2010.

\bibitem{shi2022breaking}
Y.~Shi, T.~Gao, X.~Jiao, and N.~Cao.
\newblock Breaking the fourth wall of data stories through interaction.
\newblock {\em IEEE Transactions on Visualization and Computer Graphics},
  29(1):972--982, 2022.

\bibitem{shi2021communicating}
Y.~Shi, X.~Lan, J.~Li, Z.~Li, and N.~Cao.
\newblock Communicating with motion: A design space for animated visual
  narratives in data videos.
\newblock In {\em Proceedings of the CHI Conference on Human Factors in
  Computing Systems}, pp. 1--13, 2021.

\bibitem{thoresen2016not}
J.~C. Thoresen, R.~Francelet, A.~Coltekin, K.-F. Richter, S.~I. Fabrikant, and
  C.~Sandi.
\newblock Not all anxious individuals get lost: Trait anxiety and mental
  rotation ability interact to explain performance in map-based route learning
  in men.
\newblock {\em Neurobiology of Learning and Memory}, 132:1--8, 2016.

\bibitem{blood2}
J.~Tiles.
\newblock Interview: Paul cummins - blood swept lands \& seas of red.
\newblock
  \url{https://www.johnson-tiles.com/blog/2014/11/interview-paul-cummins-blood-swept-lands-seas-red/},
  2014.
\newblock \x{Last accessed: June 25, 2023}.

\bibitem{tufte2001visual}
E.~R. Tufte.
\newblock {\em The visual display of quantitative information}.
\newblock Graphics Press, 2001.

\bibitem{valdez1994effects}
P.~Valdez and A.~Mehrabian.
\newblock Effects of color on emotions.
\newblock {\em Journal of Experimental Psychology: General}, 123(4):394--409,
  1994.

\bibitem{van2021just}
R.~van Koningsbruggen and E.~Hornecker.
\newblock “it’s just a graph”--the effect of post-hoc rationalisation on
  infovis evaluation.
\newblock In {\em Creativity and Cognition}, pp. 1--10. ACM, 2021.

\bibitem{van2010affective}
R.~van Lammeren, J.~Houtkamp, S.~Colijn, M.~Hilferink, and A.~Bouwman.
\newblock Affective appraisal of \x{{3D}} land use visualization.
\newblock {\em Computers, Environment and Urban Systems}, 34(6):465--475, 2010.

\bibitem{viegas2004artifacts}
F.~B. Vi{\'e}gas, E.~Perry, E.~Howe, and J.~Donath.
\newblock Artifacts of the presence era: Using information visualization to
  create an evocative souvenir.
\newblock In {\em IEEE InfoVis}, pp. 105--111. IEEE, 2004.

\bibitem{viegas2009participatory}
F.~B. Vi{\'e}gas, M.~Wattenberg, and J.~Feinberg.
\newblock Participatory visualization with wordle.
\newblock {\em IEEE Transactions on Visualization and Computer Graphics},
  15(6):1137--1144, 2009.

\bibitem{walter2011designing}
A.~Walter.
\newblock {\em Designing for emotion}.
\newblock A Book Apart, 2011.

\bibitem{wang2021dehumor}
X.~Wang, Y.~Ming, T.~Wu, H.~Zeng, Y.~Wang, and H.~Qu.
\newblock Dehumor: Visual analytics for decomposing humor.
\newblock {\em IEEE Transactions on Visualization and Computer Graphics},
  28(12):4609--4623, 2021.

\bibitem{wang2019emotional}
Y.~Wang, A.~Segal, R.~Klatzky, D.~F. Keefe, P.~Isenberg, J.~Hurtienne,
  E.~Hornecker, T.~Dwyer, and S.~Barrass.
\newblock An emotional response to the value of visualization.
\newblock {\em IEEE Computer Graphics and Applications}, 39(5):8--17, 2019.

\bibitem{xie2023creating}
L.~Xie, X.~Shu, J.~C. Su, Y.~Wang, S.~Chen, and H.~Qu.
\newblock Creating emordle: Animating word cloud for emotion expression.
\newblock {\em IEEE Transactions on Visualization and Computer Graphics}, 2023.

\bibitem{yang2021design}
L.~Yang, X.~Xu, X.~Lan, Z.~Liu, S.~Guo, Y.~Shi, H.~Qu, and N.~Cao.
\newblock A design space for applying the freytag's pyramid structure to data
  stories.
\newblock {\em IEEE Transactions on Visualization and Computer Graphics},
  28(1):922--932, 2021.

\bibitem{yu2017stresstree}
B.~Yu, M.~Funk, J.~Hu, and L.~Feijs.
\newblock Stresstree: A metaphorical visualization for biofeedback-assisted
  stress management.
\newblock In {\em Proceedings of the Conference on Designing Interactive
  Systems}, pp. 333--337, 2017.

\bibitem{zeller2022scientific}
S.~Zeller, F.~Samsel, and L.~Bartram.
\newblock Affective, hand-sculpted glyph forms for engaging and expressive
  scientific visualization.
\newblock In {\em IEEE VIS Arts Program}, pp. 127--136. IEEE, 2022.

\bibitem{zeng2019emoco}
H.~Zeng, X.~Wang, A.~Wu, Y.~Wang, Q.~Li, A.~Endert, and H.~Qu.
\newblock \x{{EmoCo}}: Visual analysis of emotion coherence in presentation
  videos.
\newblock {\em IEEE Transactions on Visualization and Computer Graphics},
  26(1):927--937, 2019.

\bibitem{zhang2010affective}
S.~Zhang, Q.~Huang, S.~Jiang, W.~Gao, and Q.~Tian.
\newblock Affective visualization and retrieval for music video.
\newblock {\em IEEE Transactions on Multimedia}, 12(6):510--522, 2010.

\end{thebibliography}


\end{document}